\documentclass[aip]{revtex4-1}
\usepackage{url} 
\usepackage{amsmath,amssymb}
\usepackage{soul}
\usepackage{bm}
\usepackage{color}
\usepackage{graphicx}

\begin{document}

\newcommand{\change}[1]{{ #1}}

%
%

\title{Nonlinear coupling of whistler waves to oblique electrostatic turbulence enabled by cold plasma}

%
%




\author{Vadim Roytershteyn}
\email[Corresponding author. Email: ]{vroytershteyn@spacescience.org}
\affiliation{ 
Space Science Institute, Boulder, CO 80301, USA
}%
\author{Gian Luca Delzanno}
\affiliation{T-5 Applied Mathematics and Plasma Physics Group, Los Alamos National Laboratory, Los Alamos, NM 87545, USA%
}%




\begin{abstract}
Kinetic simulations and theory demonstrate that whistler waves can excite oblique, short-wavelength fluctuations through secondary drift instabilities if a population of sufficiently cold plasma is present.
 The excited modes lead to heating of the cold populations and damping of the primary whistler waves. The instability threshold depends on the density and temperature of the cold population and can be relatively small if the temperature of the cold population is sufficiently low. \change{This} mechanism may thus play a significant role in controlling amplitude of whistlers in the regions of the Earth's magnetosphere where cold background plasma of sufficient density is present. 
\end{abstract}

\maketitle

\section{Introduction}

Whistler waves are electromagnetic plasma modes with frequency between the ion and the electron cyclotron frequencies. Famous for their characteristic dispersion relation, whistler waves are frequently observed in the solar wind and are ubiquitous in the Earth's magnetosphere~\cite{helliwell69}.  They play a major role in the dynamics of the latter~\cite{thorne2010}, where they appear either as "chorus", discrete emissions typically in two distinct bands \cite{burtis69,tsurutani74,santolik09,li2013}, or "hiss",  broadband emissions found predominantly in the plasmasphere ~\cite{dunckel69,russell69,thorne73,hartley18} and in plasmaspheric plumes\cite{Chan1976}. The chorus waves are associated with local energization of energetic particles~\cite{summers1998,meredith02,meredith03}, as well as electron precipitation in the form of diffuse~\cite{Ni14} and pulsating aurora~\cite{nishimura10,kasahara2018}, or microbursts~\cite{oliven68,breneman17}. Whistler waves observed in the Earth's magnetosphere can reach very large amplitudes, e.g.~\cite{Cattell2008,Wilson2011,Tyler2019a,Tyler2019b}. 

Naturally occurring chorus waves in the Earth's magnetosphere are primarily generated by an instability driven by temperature anisotropy of hot ($\sim$keV) electrons~\cite{kennel66,Gary:1993}, which are injected into the magnetosphere during substorms. This whistler instability evolves to reduce the temperature anisotropy, driving the electron distribution towards a marginally stable state, and thus providing an upper bound for the value of the hot electron temperature anisotropy~\cite{gary96,xiao06,tao17,an17,  gary_grl05, macdonald08, yue16}. Similar processes have been identified in the solar wind~\cite{Stverak2008}.  In addition to hot electrons driving the instability, cold plasma populations are commonly found in the magnetosphere. They generally originate from \change{the} ionosphere and in many regions dominate the total plasma density. It is well appreciated that such populations may affect the growth rates and the saturation level of whistler instability~\cite{cuperman74,gary12,wu13,cuperman73}. However, cold plasma is generally thought of as a "passive" player, simply providing the inertia of the medium. 

In this work, we use kinetic simulations and theory to propose a new scenario in which, in the presence of cold electron populations, whistler waves of sufficiently large amplitude can excite oblique, short-wavelength fluctuations through drift-type secondary instabilities leading to heating of the cold populations and damping of the primary whistler waves. Despite its potential significance, the particular coupling discussed here does not appear to have been previously identified in the literature, although
several conceptually similar processes have been discussed. For example, Khazanov et al.\cite{Khazanov2017} reported observations of lower-hybrid oscillations associated with electromagnetic ion cyclotron (EMIC) waves and attributed them to an instability excited by differential drifts between ions and electrons driven by the electric field of EMIC waves. Saito et al.\cite{Saito2015} demonstrated that large amplitude magnetosonic-whistler modes with frequencies of the order of proton cyclotron frequency and wavelegnths of the order of several proton inertial lengths  can excite the Modified Two Stream Instability (MTSI), which leads to damping of the primary mode. Similarly to these previous studies, the instabilities reported here are driven by the differential drifts between plasma components. However, they are distinct in that they involve primarily the cold populations, involve coupling with different plasma modes, and operate on different time scales compared to the studies of Refs.~\cite{Khazanov2017} and~\cite{Saito2015}. The latter is an important distinction, since only relatively fast instabilities can affect short-wavelength whistlers whose frequency generally exceeds the lower-hybrid frequency.

We note that in general, a multitude of nonlinear processes associated with whistler waves have been previously identified and extensively studied, in part due to their significance for magnetospheric dynamics. Some examples include nonlinear wave-particle interactions that are thought to be responsible for generation of chorus rising or falling tones~\cite{Omura2012}, parametric interaction of whistler waves with electrostatic modes~\cite{Boswell1977,Umeda2017}, 3-wave coupling processes~\cite{Fu2017}, nonlinear scattering~\cite{Ganguli2010}, and several others, e.g. ~\cite{Artemyev2016,Demekhov2017,Agapitov2018,Vasko2018}. It is interesting to note that the mechanism discussed here has an amplitude threshold that depends on density and temperature of the cold plasma population. While these parameters are often not known accurately, this opens the possibility that the threshold can be comparable or lower than that of many of the previously identified processes.

The process discussed here may thus have important implications for the Earth's magnetosphere, and possibly other systems, 
since it may control the amplitude of the whistler waves and hence the rate of pitch-angle scattering and energization of systems like plasma sheet, ring current and radiation belts.


\section{Methods}
To illustrate the essential physics of the process, we focus on whistler waves generated by the whistler anisotropy instability and consider a simple local model. We note, however, that the secondary instabilities discussed below do not depend on the particular process generating the primary whistler waves. We performed particle-in-cell (PIC) simulations using the VPIC code~\cite{Bowers2008}, which solves the system of relativistic Vlasov-Maxwell equations. The initial conditions correspond to three (bi-)Maxwellian particle populations: cold isotropic electrons with density $n^C$ and temperature $T_{e0}^C$, hot anisotropic electrons with density $n^{H}$, parallel temperature $T_{e0,\parallel}^H$, and perpendicular temperature $T_{e0,\perp}^H$, and cold isotropic ions with temperature $T_{i0}=T_{e0}^C$ and density $n_0=n^C+n^H$. A uniform magnetic field $B_0$ oriented in the $z$-direction was imposed at time $t=0$ and the parallel and perpendicular directions are defined with respect to $B_0$. In the present discussion, we focus predominantly on 2D cases distinguished by the level of anisotropy of the hot electron population $A^H=T_{e0,\perp}^H/T_{e0,\parallel}^H-1$, which is the driver of the primary instability. Consequently, the amplitude of the excited whistler waves differs between the simulations. For each case we perform one- and two-dimensional (1D/2D) simulations. Additionally, results from a  3D simulation are used to illustrate that when the conditions are favorable, the processes under consideration may lead to almost complete damping of the primary whistler waves.

The parameters of the high-anisotropy case correspond to $T_{e0}^C=10$ eV, $n^C/n^H=4$, $T_{e0,\|}^H/T_{e0}^C=200$ and $A^H=4$. The parallel electron beta for the hot electrons is $\beta_{||e}^H=8\pi n_H T_{e0,\|}^H/B_0^2=2.5\times 10^{-2}$, while the cold electrons have $\beta_{e}^C=5\times 10^{-4}$. For these parameters, the most unstable modes of the whistler instability are field-aligned~\cite{gary12} and the maximum growth rate corresponds to parallel wavenumbers $k d_e \sim 1$. Here $d_s = c/\omega_{ps}$ is the reference inertial length for species $s$ (ions or electrons) with mass $m_s$, and $\omega_{ps}^2= 4 \pi n_0 e^2/m_s$. The 2D simulation has domain size $L_y \times L_z = (0.4\pi \times 20 \pi) d_e$ with $ n_y \times n_z = 304 \times 15200$ cells. The minimum allowed parallel wavenumber is thus $k_\parallel^{\mathrm{min}} d_e = 0.1$, which allows several modes with growth rates near the peak to grow. The average number of particles per cell per particle species (ions and electrons) is $N_\mathrm{ppc} = 10^4$. The time step is $\Delta t\omega_{pe} \approx 0.0029$. The ratio of the reference plasma frequency to the electron cyclotron frequency  is $\omega_{pe}/\Omega_{ce}=4$, where $\Omega_{cs} = e B_0/(m_s c)$. The chosen parameters are consistent with geomagnetically-active conditions measured at geosynchronous orbit by the Los Alamos National Laboratory Magnetospheric
Plasma Analyzer (MPA) instruments~\cite{joe2013}.
They are also consistent with the parameters used by Yu {\em et al.}\cite{yu2018}, which were  obtained from a ring-current/plasmaspheric model.
The parameters of the low anisotropy case are $T_{e0}^C=1$ eV, $n^C/n^H=4$, $T_{e0,\|}^H/T_{e0}^C=2000$ and $A^H=2$, corresponding to  $\beta_{||e}^H=2.5\times 10^{-2}$ and $\beta_{e}^C=5\times 10^{-5}$, and again the most unstable modes are field-aligned. The other parameters are $L_y \times L_z = (0.2\pi \times 2 \pi) d_e$, $ n_y \times n_z = 540 \times 5000$ cells,  $N_\mathrm{ppc} = 10^4$, and $\Delta t\omega_{pe} \approx 8.4 \times 10^{-4}$. The 1D simulations for each case have identical parameters to the corresponding 2D ones and the computational domain is along $B_0$. The simulations described here properly resolve spatial and temporal scales associated with the cold population (such as the Debye length) and, as a consequence, are computationally challenging. 

Additionally, we have performed a 3D simulation with parameters corresponding to the high-anisotropy case. Due to a high computational cost of such simulations, we reduced the size of the domain to $L_z \approx 5.5 d_e$ and $L_x = L_y = 0.2 \pi d_e $, such that only a single primary whistler mode can be accommodated in the simulation domain. The resolution of the domain was also reduced to $n_x=n_y=48$ and $n_z=512$ cells. The corresponding time step was $\Delta t \omega_{pe} \approx 0.007$, while other parameters, such as the number of particles per cell, remained the same as in the 2D case.

\section{Results}
\begin{figure}[h]
\centering
\includegraphics[width=4in]{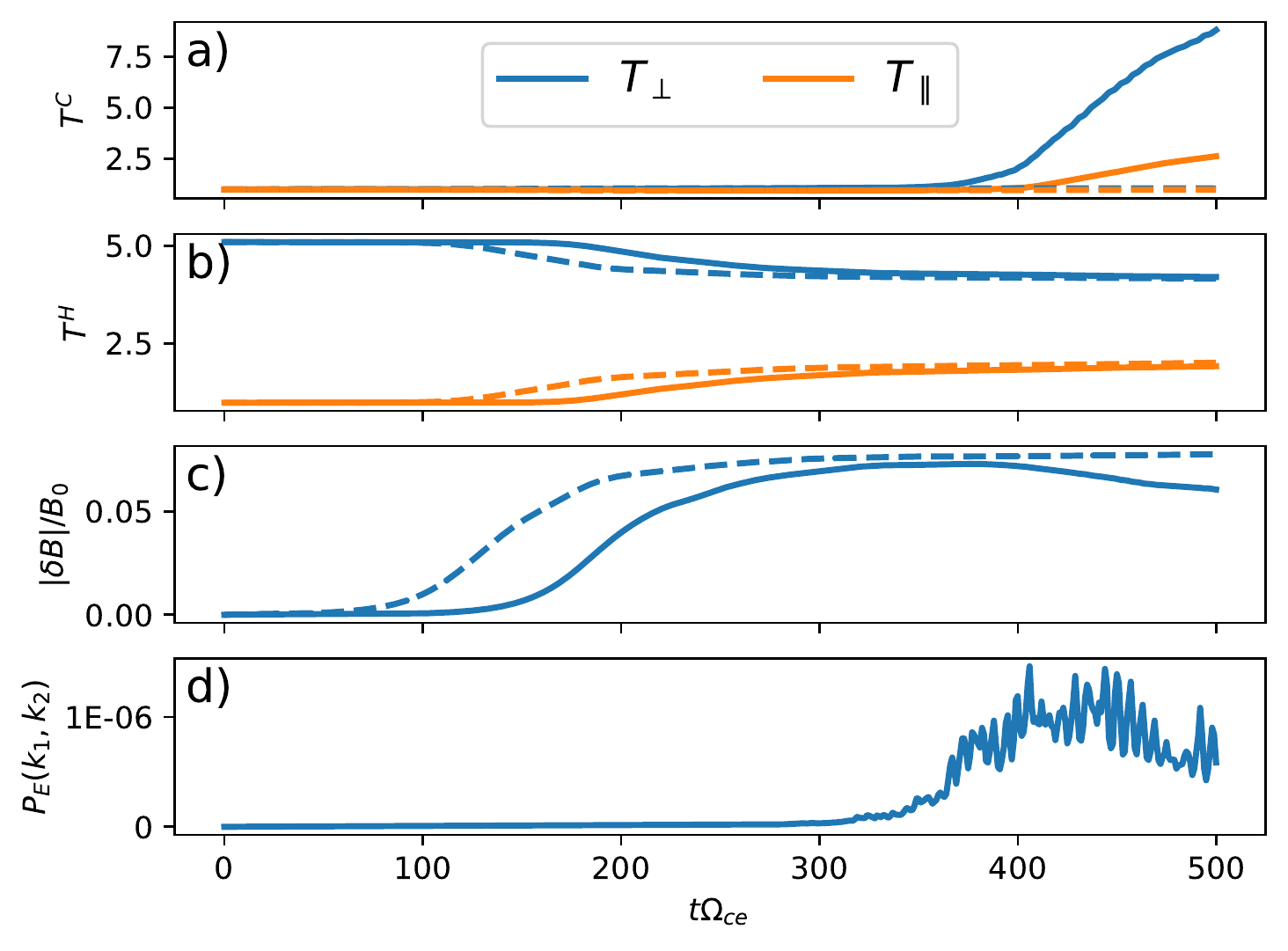}
\caption{Time evolution of various quantities in the high-anisotropy simulation: (a) the parallel and perpendicular temperatures of the cold electron population, (b) the parallel and perpendicular temperatures of the hot electron population, (c) the amplitude of magnetic field fluctuations, and (d) the energy associated with finite-$k_y$ fluctuations of the transverse electric field $E_y$ in the range of wavenumbers between $k_1 \rho_e^C\approx 0.25$ and $k_2 \rho_e^C \approx 19$. The solid (dashed) lines correspond 2D (1D) simulations.}
\label{fig:mainresult5}
\end{figure}

Fig.~\ref{fig:mainresult5} summarizes the important features of \change{the} high-anisotropy simulation. Panel a) shows evolution of the parallel and perpendicular temperatures of the cold electrons in 2D and 1D simulation with $A^H=4$, while panel b) shows evolution of the hot electron temperature. The  imposed anisotropy of the hot electrons leads to development of the whistler anisotropy instability, which we will refer to as the "primary" instability. The instability leads to growth of magnetic fluctuations $\delta B$, as shown in panel c), and partial isotropization of the hot population in the time interval $ 125 \lesssim t\Omega_{ce} \lesssim 250$. This is a well-known result~\cite{gary96}. Note that in the simulations the instability grows out of numerical noise and since the noise properties differ in 1D and 2D simulations, the respective time traces are shifted in time. Because the primary instability is non-resonant with the cold population, its development does not have an appreciable effect on the temperatures of the latter. However, at later times, the cold electrons experience strong perpendicular and somewhat weaker, but still appreciable, parallel heating.  The heating is only present in the 2D case and is associated with the development of short-wavelength,  oblique electrostatic turbulence. This is illustrated in panel d) of Fig.~\ref{fig:mainresult5}, which shows the power in small-scale electric field fluctuations $P_E(k_1,k_2) = \sum_{|k_y|=k_1}^{k_2} \langle |\hat E_{y}(k_y,z)|^2 \rangle_z$, where $\hat E_y$ refers to the Fourier transform (FT) of $E_y$, and  $\langle \cdot \rangle_z$ is the spatial average over $z$. The short-wavelength fluctuations grow after the saturation of the primary instability at around $t\Omega_{ce} \sim 400$. Their growth is correlated with a decrease in the amplitude of the magnetic fluctuations and heating of the cold population. Also, it is evident from Fig. \ref{fig:mainresult5} that cold plasma heating occurs after the electrostatic fluctuations have reached sufficient amplitude $t\Omega_{ce} \gtrsim 375$. As will be shown below, the electrostatic turbulence is due to the development of a secondary instability with relatively high $k_\perp$, which is the reason why it cannot be captured in 1D configurations. 

\begin{figure}[h]
\centering
\includegraphics[width=3in]{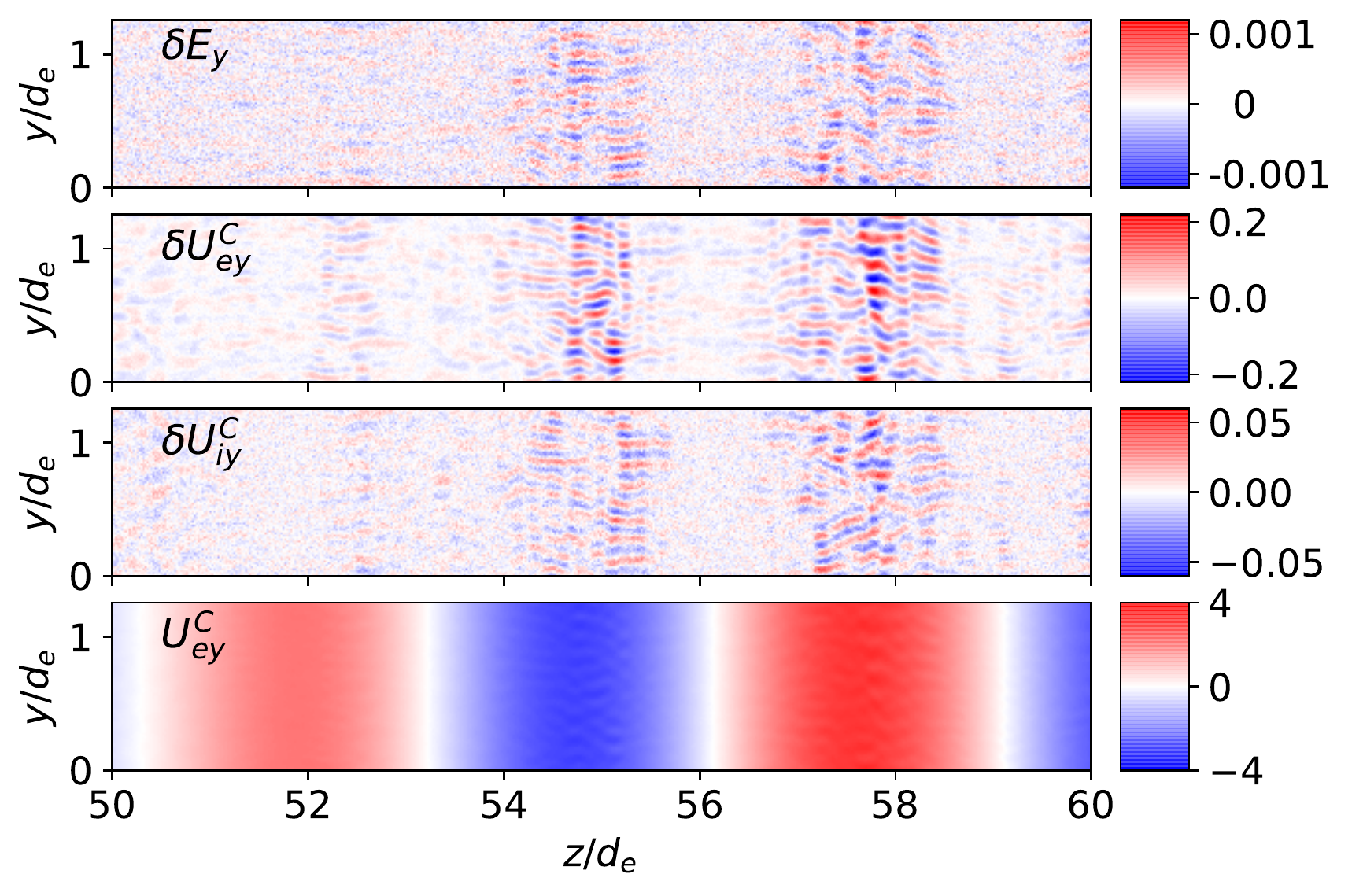}
\caption{Spatial profiles of (top to bottom) electric field fluctuations $\delta E_y$, velocity fluctuations of cold electrons $\delta U_{ey}^C$ and  ions $\delta U_{iy}$ in the high-anisotropy simulation at $t \Omega_{ce} = 300$. For each quantity, the plots show deviation from the $y$-averaged value, e.g. $\delta U = U - (1/L_y) \int_0^{L_y} U d y  $. The bottom panel shows full profile of $U_{ey}^C$. The velocities of electrons and ions are normalized to the initial thermal velocity of the  population. Only a small part of the simulation domain is shown.}
\label{fig:multi}
\end{figure}

\begin{figure}[h]
\centering
\includegraphics[width=3in]{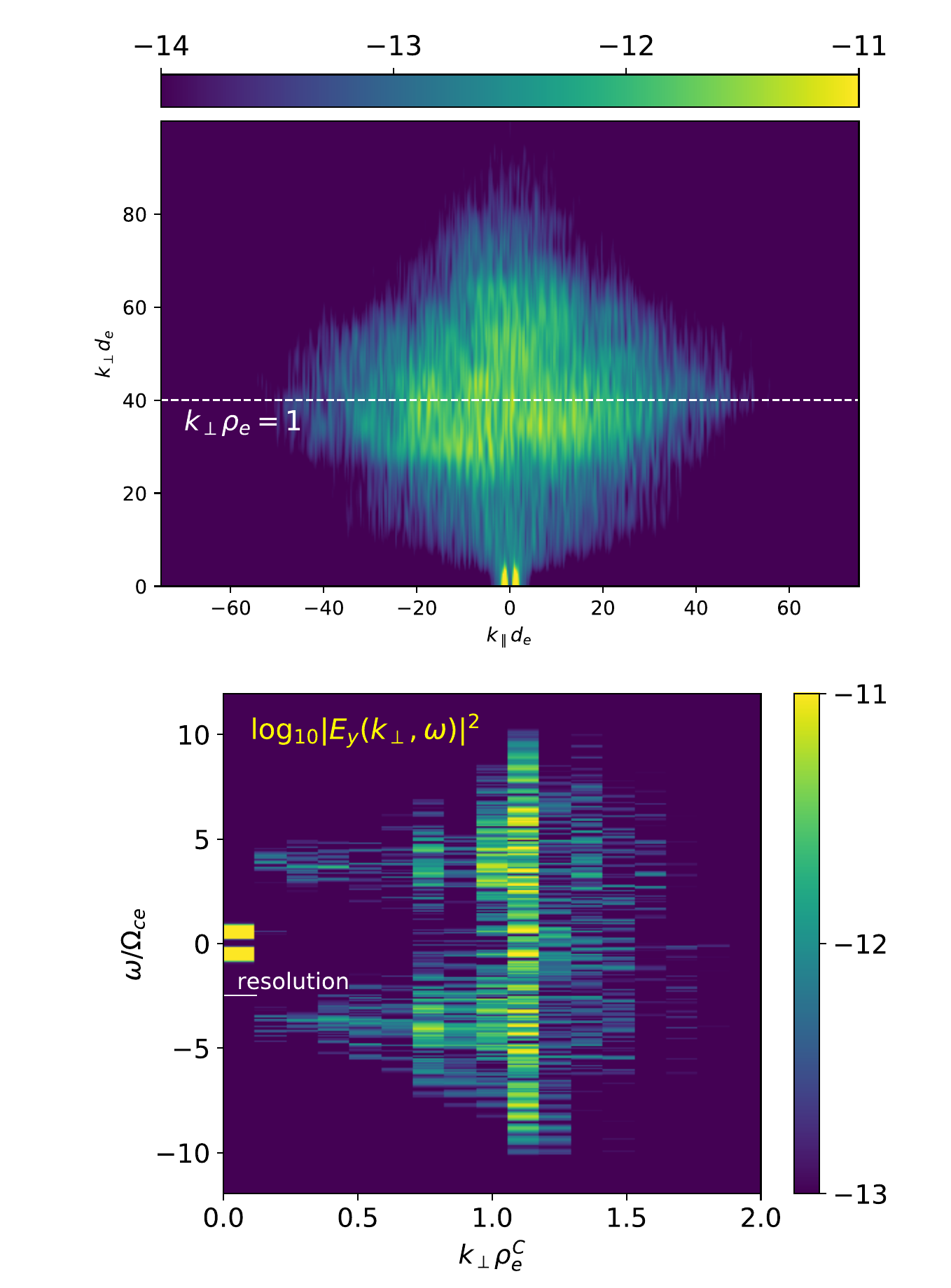}
\caption{Spectral characteristics of $E_y$ fluctuations in the high-anisotropy simulation. Top: $k_\perp$ -- $k_\parallel$ spectrum at $t \Omega_{ce} = 400$. Bottom: frequency -- $k_\perp$ spectrum obtained by performing Fast Fourier Transform of data collected at $z\approx 0.08 d_e$ during $t \Omega_{ce} = 250-375$.}
\label{fig:eyspectrum}
\end{figure}

Spatial profiles of $E_y$ fluctuations, as well as the velocity fluctuations of the cold electrons and ions are shown in Fig.~\ref{fig:multi}. We observe that the short-wavelength fluctuations \change{couple cold ions and cold electrons and} appear to be associated with the transverse electron flow driven by the primary whistler waves. The top panel in Fig.~\ref{fig:eyspectrum} shows wavenumber spectrum $|\hat E_y(k_y,k_z)|^2$ at $t \Omega_{ce} = 400$. The fluctuations are broadband, short-wavelength, and are excited in a range of angles with respect to $B_0$. The bottom panel illustrates frequency spectrum, obtained by performing FT in time and $y$ of electric field $E_y(y,z_0,t)$ collected at a fixed $z=z_0\approx 0.08d_e$. For the chosen time interval, the dominant fluctuations are  characterized by $k_\perp \rho_e^C \lesssim 1$ and $k_\parallel \lesssim k_\perp$. Here $\rho_e^C = v_{te\perp}^C/\Omega_{ce}$ is the gyroradius of the cold electrons defined with $v_{te\perp}^C = (2 T_{e\perp}^C/m_e)^{1/2}$. For the parameters of the high-anisotropy simulation the reference value is $\rho_e^C/d_e = \sqrt{(n_e/n_e^c)\beta_e^C} = 0.025$. The fluctuations are predominantly electrostatic, with most power in transverse $E_y$ fluctuations at the time shown, although significant signal is also present in the $E_z$ component (see also discussion below). Here and elsewhere in the paper the electric field is normalized to the value $E^* = m_e c \omega_{pe}/ e$. 

\begin{figure}[h]
\centering
\includegraphics[width=4in]{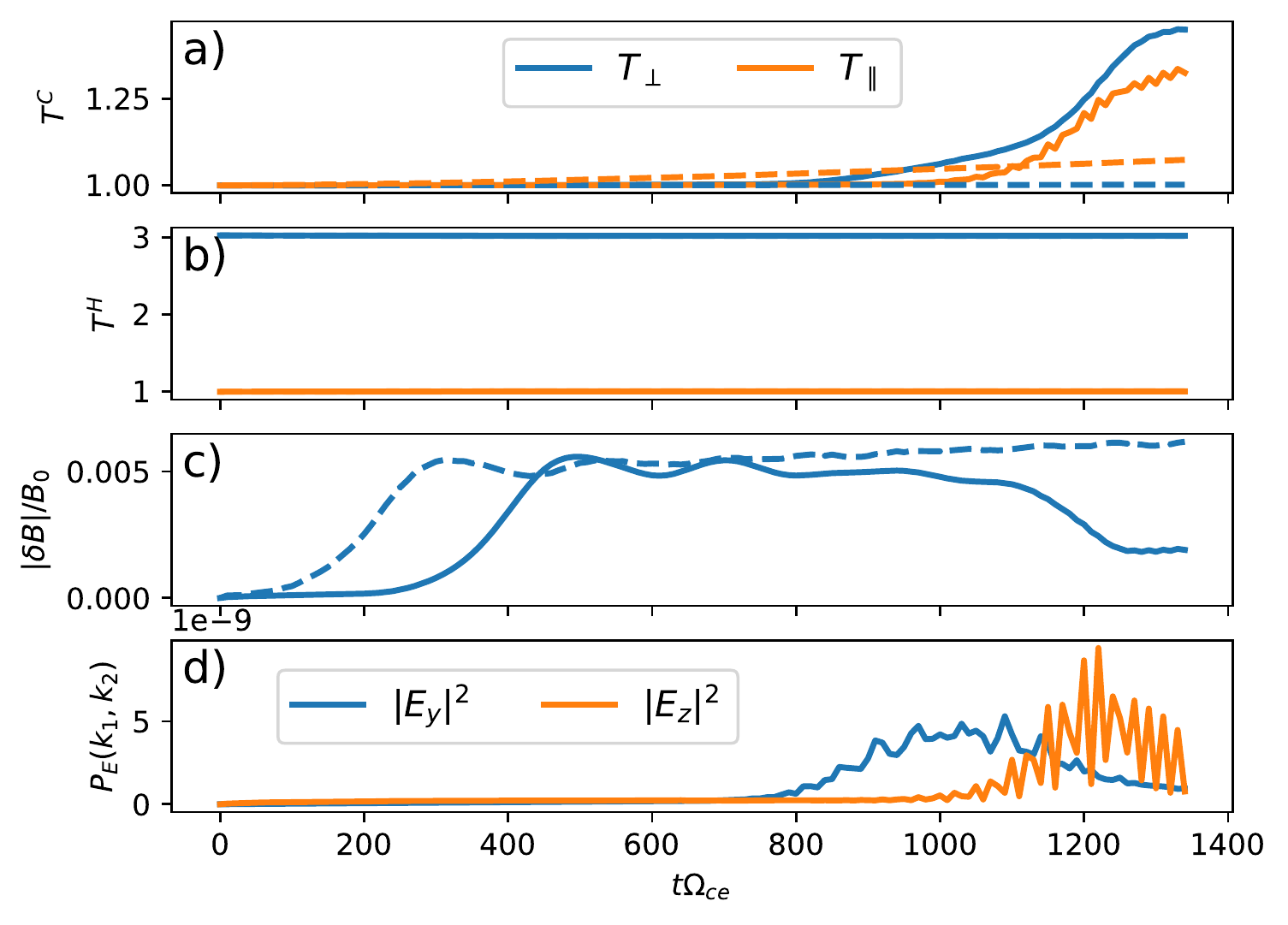}
\caption{Time evolution of various quantities in the low-anisotropy simulation: (a) the parallel and perpendicular temperatures of the cold electron population, (b) hot electron temperature, (c) amplitude of magnetic field fluctuations, and (d) the energy associated with finite-$k_y$ fluctuations of the transverse $E_y$ and parallel $E_z$ electric field components. The $E_y$ energy is computed in the range of wavenumbers between $k_1 \rho_e^C \approx 1$ and $k_2 \rho_e^C \approx 2.6$, while for $E_z$ power  $k_1 \rho_e^C \approx 0.08$ (minimum allowed by the simulation domain) and $k_2 \rho_e^C \approx 1$. In panels a) -- c) the dashed lines show the same quantities in the corresponding 1D simulation.}
\label{fig:mainresult3}
\end{figure}

It is important to emphasize that the secondary instabilities can develop at any amplitude of the primary mode, provided that the background electrons are sufficiently cold. This is illustrated by Fig.~\ref{fig:mainresult3}, which shows results from the  simulation with $A^H = 2$ in the same format as Fig.~\ref{fig:mainresult5}. Overall, the dynamics resemble the higher-$A^H$ case, cf. Fig.~\ref{fig:mainresult5}. However, the primary instability now saturates at much lower amplitude, $\delta B/B_0 \sim 5 \times 10^{-3}$. Since the amplitude of the fluctuations and the change in the perpendicular temperature of the hot population are related by $ \delta B^2/B_0^2 \propto \beta_{\perp e}^H \delta (T_{\perp e}^H / T_{\perp e,0}^H )$, the hot population experiences much weaker change in its temperatures, such that $\delta T_{\perp e}^H/T_{\perp e,0}^H \sim 10^{-3}$. Despite relatively small $\delta B$,  a wide variety of secondary instabilities still develops and leads to decay of the primary mode and a modest heating of the cold electron population. The magnetic field fluctuations saturate to a finite value $\delta B/B_0 \sim 1 \times 10^{-3}$ after the development of the secondary instabilities. Also notable in Fig.~\ref{fig:mainresult3} is the fact that the most intense damping of the primary whistler wave is associated with growth of oblique short-wavelength $E_z$ fluctuations, as seen in Panel d).  Note that the cold electron population has $\beta_e^C = 5 \times 10^{-5}$ in this case, compared to $\beta_e^C = 5 \times 10^{-4}$ in the high-anisotropy case.

To identify the nature of the secondary instabilities, we recall that large transverse drifts between electrons and ions can drive a variety of instabilities (\change{see e.g. a review in Ref}.~\cite{Muschietti2017} and the references therein). In the cases discussed here, the drifts are due to fluctuating current of the primary whistler waves and can be significant in relation to the thermal speed of the cold electron component, provided its temperature is low enough: $V_d\change{/v_e^C } \sim j/n_0 e v_e^C \sim (k_\| d_e) (\delta B/B_0)/\sqrt{\beta_e^C}$. 

A simple model could be obtained by focusing on electrostatic modes. We treat the primary whistler mode as a given driver with electric field $\bm E_D(t) = \bm E_0 e^{ i \omega_0 t }$, where $\omega_0 \lesssim \Omega_{ce}$ is the driver frequency. For the parameters considered in the simulations $\omega_0 \approx 0.5 \Omega_{ce}$.
Since we ignore spatial variation of the driving field, the analysis below is applicable to relatively short-wavelength fluctuations, such that $k_z \gg k_z^0$ where $k_z^0$ is the wavenumber associated with the primary whistler. The cold electrons respond to $\bm E_D$ with drifts $\bm V_c(t)$ that satisfy $m_e d \bm V_c/dt = -e \bm E_D - (e/c) {\bm V}_c \times {\bm B_0}$. A transformation into the frame of reference co-moving with the cold electron population can be obtained by a change of coordinates $\bm x' = \bm x - \int^t \bm V_c d \tau$ and $\bm v' = \bm v  - {\bm V}_c$. Assuming electrostatic perturbations, the linearized Vlasov equation for the electrons in this co-moving frame takes the usual form\change{~\cite{Kaw1973}}
\begin{equation}
    \partial_t f_e + \bm v' \cdot \nabla_{\bm x'} f_e - \frac{e}{m_e} \left [ \tilde {\bm E} + (\bm v' /c) \times {\bm B}_0  \right ] \cdot \nabla_{\bm v'} f_e = 0,
\end{equation}
where $\tilde {\bm E} = \change{\bm E - \bm E_D}$ is the \change{perturbed} electric field. The response of cold electrons in the co-moving frame can be easily evaluated~\cite{Stix1992} as 
\begin{equation}
    \tilde n_e^C = \tilde \phi(\omega,k) \frac{ e n_e^C  }
    {T_e^C}\left [  1 + \sum_n \frac{\omega}{k_\parallel v_{te}} e^{-\lambda} I_n(\lambda) Z(\xi_n)\right]
    \label{eq:electrons_oblique}
\end{equation}
where $\tilde \phi(\omega,k)$ is the Fourier component of the electrostatic potential in the co-moving frame, $\xi_n = (\omega - n \Omega_{ce})/(k_\parallel v_{te}^C)$, $\lambda = k_\perp^2 T_e^C/(m_e \Omega_{ce}^2)$,  $Z$ is the plasma dispersion function~\cite{fried61},   and $I_n$ is the modified Bessel function of the first kind.  We assume that the ion response is unmagnetized in the relevant range of frequencies and evaluate it in the rest frame of the simulation, since ion drift due to the primary whistler field is negligible: 
\begin{equation}
\hat n_i = \frac{e n_{i} }{2 T_i} Z'\left( \frac{\omega}{k v_{ti}} \right)  \hat \phi 
\label{eq:ion_response}
\end{equation}
where $Z'=dZ(\xi)/d\xi = - 2[ 1+\xi Z(\xi)] $,  $v_{ti} = \sqrt{2 T_i/m_i}$, $k = (k_\perp^2 + k_\parallel^2)^{1/2}$, and $\hat{}$ is used to denote Fourier components in the rest frame.

Using the formalism of Kaw and Lee\cite{Kaw1973} (see also Ref.~\cite{Gamayunov1992}), we can relate the Fourier components of any quantity $A$ in the co-moving frame $\tilde A$ to those in the stationary (ion) frame $\hat A$ as
$ \tilde A (\omega, k) = \sum_m J_m(a) \hat A(\omega + m \omega_0 , k )$, 
where $J_m(a)$ is the Bessel function of argument $a = k_y |V_c|/\omega_0$. For example, the ion response transformed into the co-moving frame and expressed through Fourier components of the electrostatic potential in that frame is
\begin{align}
\tilde n_i (\omega, k) &=  \frac{e n_{0} }{ v_{ti}^2 m_i} \sum_{m'}  \sum_m J_m(a)  J_{m'}(a) \\
& Z'\left( \frac{\omega + m \omega_0}{k v_{ti}} \right)   \tilde \phi(\omega +m \omega_0 - m' \omega_0 , k ).
\label{eq:ion_response_transformed}
\end{align}
The dispersion relation follows from the Poisson's equation $ k^2 \tilde \phi (\omega, k)  = 4\pi e (\tilde n_i - \tilde n_e)$ and couples perturbations at frequencies separated by the  harmonics of $\omega_0$. The full dispersion relation can be solved numerically by considering a finite number of sidebands around a given frequency $\omega$ and numerically finding the value $\omega$ that minimizes the determinant of the resulting matrix. Simplified equations can also be obtained in  limiting cases of interest, as described below. 

First, we consider perturbations perpendicular to $B_0$. If the relative drift between cold electrons and ions were constant and equal to the peak value observed in the simulation, the relevant instability with peak growth rate in the range of wavenumbers observed in the simulations would be electron-cyclotron-drift instability (ECDI)~\cite{Forslund1970}. Under the conditions of the simulations and with constant drifts, classical dispersion relation for ECDI~\cite{Forslund1970} predicts growth rate $\gamma_\mathrm{ECDI}/\omega_{ce} \sim 0.2-0.3$ in the vicinity of $k \rho_e^C \sim 1$. However, because $\gamma_\mathrm{ECDI}$ is comparable to the typical frequency of the driver (primary whistler mode) $\omega_0$, a more complex analysis is required, taking into account the oscillations of the relative drifts in response to the driving electric field of the primary mode.

\begin{figure}[h]
\centering
\includegraphics[width=5in]{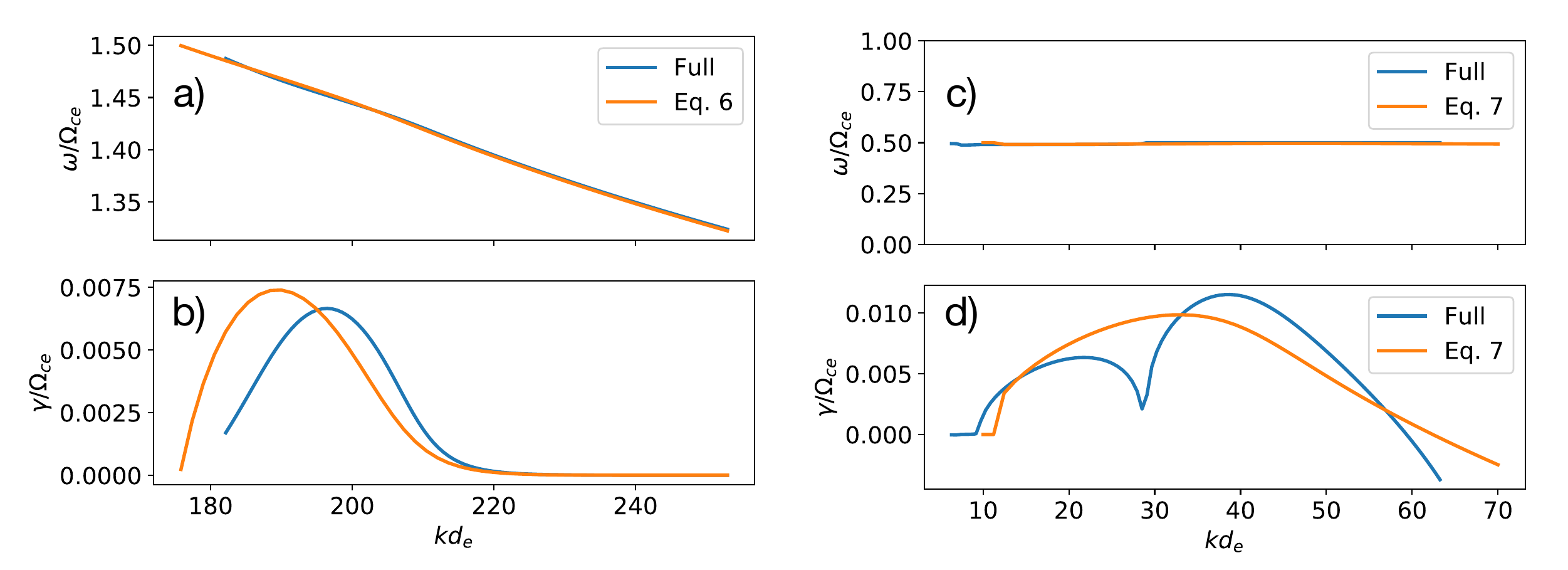}
\caption{Solutions of the linear dispersion relation for secondary instabilities for parameters of the low-anisotropy case: Panels a) and b) show frequency and the growth rate obtained by solving  Eq.~(\ref{eq:disp_simple}) and a full dispersion relation for the most unstable mode for $V_c \approx 0.65 v_{the}^C$. The solution of the full dispersion relation is obtained by considering five sidebands on each side of a given frequency. Panels c) and d) show frequency and the growth rate for the oblique mode at $\theta=60^\circ$ obtained by solving Eq.~\ref{eq:electrons_oblique_simple} and the full dispersion relation.}
\label{fig:disp}
\end{figure}

 A simple form of the dispersion relation can be obtained by taking an appropriate limit~\cite{Stix1992}  $\theta \to 90^\circ$ of Eq.~\ref{eq:electrons_oblique} and observing that for the parameters considered $ |\omega_0 /(k v_{ti}) | \gg 1$, so that the ion response can be ignored, except  near frequencies satisfying $|\omega + m^* \omega_0| \sim k v_{ti} $ for some integer value of $m^*$. Retaining only the electron terms yields the usual dispersion relation for the electron Bernstein (EB) modes~\cite{Stix1992}. Instabilities potentially appear near the intersection of the EB modes with the harmonics of $\omega_0$. Writing $\omega = -m^* \omega_{0} + \delta \omega$,  we obtain 
\begin{equation}
 1 -  \frac{n^C}{n_0}  \frac{ 2  \omega_{pe}^2}{\lambda} \sum_{n=1}^{\infty} \frac{e^{-\lambda} I_n(\lambda) n^2 }{\omega^2 - n^2 \Omega_{ce}^2}  =  \frac{\omega_{pi}^2 }{k^2 v_{ti}^2} \left [J_{m^*}(a) \right ]^2 Z' \left ( \frac{\delta \omega}{k v_{ti}} \right ). 
 \label{eq:disp_simple} 
\end{equation}
Here we assumed that for a given $k$ there exists only one intersection of the EB dispersion relation with the harmonics of $\omega_0$ and ignored couplings to the other modes. Fig.~\ref{fig:disp} shows an example of the solution of Eq.~(\ref{eq:disp_simple}) for the parameters corresponding to the low-anisotropy case \change{and compares it with a solution of the full dispersion relation retaining modes at several harmonics of $\omega_0$}. Panel a) and b) show the frequency and the growth rate as a function of wavenumber for $V_c = 0.65 v_{te}^C$. The branch shown in panels a) and b) corresponds to the intersection of the first EB mode with the third harmonic of the primary whistler mode (such that $m^* = -3$).

The quasi-perpendicular modes \change{associated with ECDI} appear first in the simulations \change{and could be easily identified by the increase in $E_y$ fluctuations and associated perpendicular heating of the cold populations in Fig.~\ref{fig:mainresult5} and~\ref{fig:mainresult3}}. However, the most significant energy transfer between the primary whistler and the cold plasma appears to be associated with development of \change{a distinct class of} oblique modes \change{that are responsible for the increase in $E_z$ fluctuations in Fig.~\ref{fig:mainresult5} and~\ref{fig:mainresult3}}. We note that some properties of the oblique instabilities can be deduced by taking the limit corresponding to the cold plasma approximation $\lambda \to 0$ and $\xi \gg 1 $ in Eq.~\ref{eq:electrons_oblique}.  In this limit, the modes of interest are electrostatic whistlers in the co-moving frame with wavevectors near the critical angle $\cos \theta_c \change{\approx}  \omega /\Omega_{ce}$, where $\omega$ is the frequency in the co-moving frame. We observe that potential instabilities arise at the intersection of electrostatic whistler dispersion relation with Doppler-shifted ion response, i.e. $|\omega - \omega_0| \sim k v_{ti} \ll \omega_0$. 

Using the same arguments as for the perpendicular modes, we arrive at the corresponding dispersion relation valid in the vicinity of the whistler branch
\begin{equation}
    k^2 + 
    \frac{4 \pi n_e^C e^2}{T_e^C}\left [  1 + \sum_n \frac{\omega}{k_\parallel v_{te}} e^{-\lambda} I_n(\lambda) Z(\xi_n)\right] = \frac{\omega_{pi}^2 }{v_{ti}^2} \left [J_{1}(a) \right ]^2 Z' \left ( \frac{ \omega - \omega_0}{k v_{ti}} \right ).
    \label{eq:electrons_oblique_simple}
\end{equation}
In practice, it is sufficient to keep only a few terms in the sum over the Bessel functions. An instability could also be found when a cold limit is taken for the electron terms on the left-hand side of Eq.~\ref{eq:electrons_oblique_simple}, but such an analysis significantly overestimates the growth rate and does not yield correct behavior at large $k$. An example of the solution of Eq.~\ref{eq:electrons_oblique_simple} for parameters relevant to the low-anisotropy simulation is shown in Panels c) and d) of Fig.~\ref{fig:disp}, where the numerical solution of the full dispersion relation coupling three sidebands is also shown. 

\begin{figure}[h]
\centering
\includegraphics[width=4in]{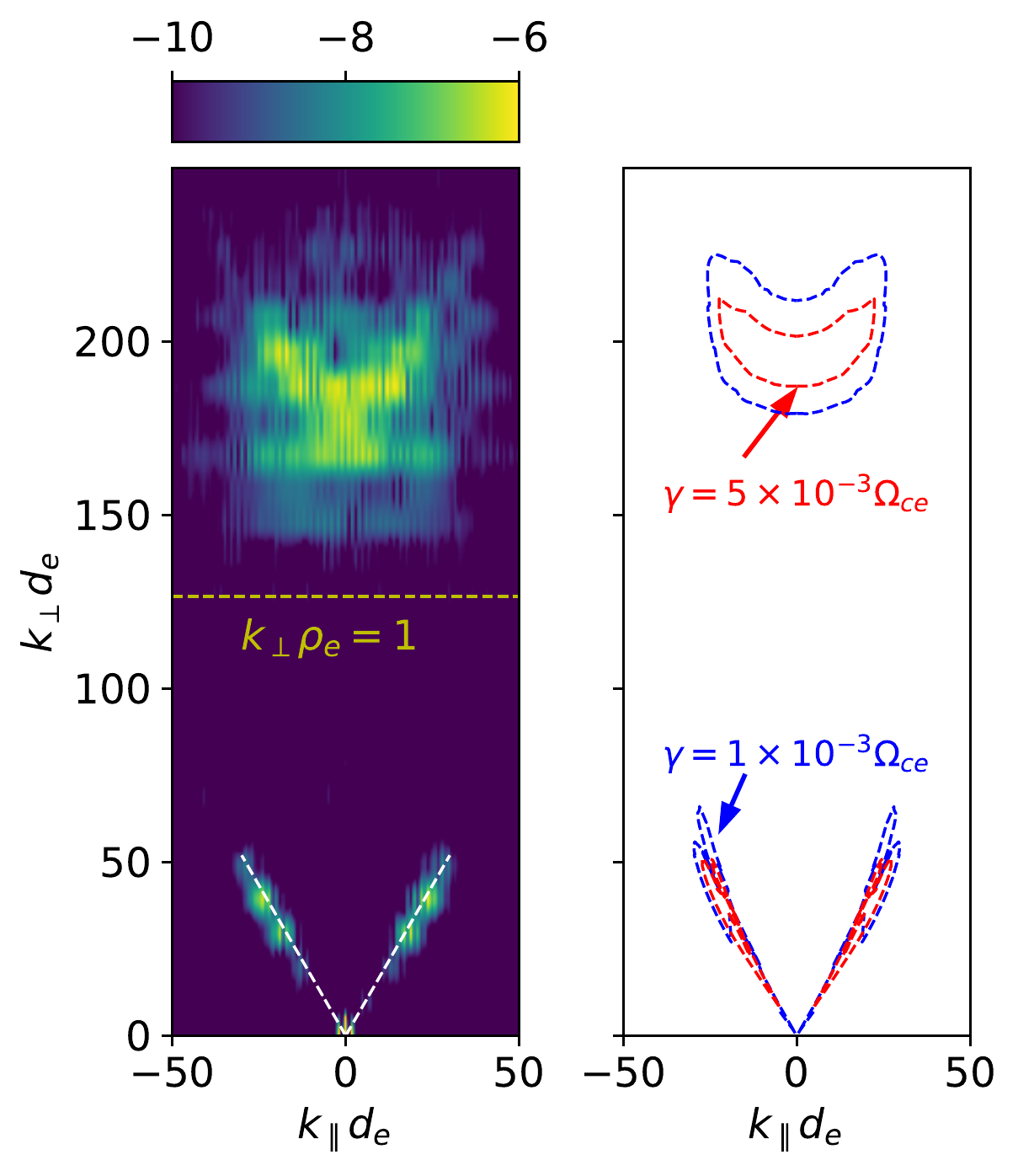}
\caption{Left: $k_\perp - k_\parallel$ spectrum of ion density fluctuations in the low-anisotropy simulation at time $t\Omega_{ce} \approx 900$. The white dashed lines correspond to the resonance cone $\cos \theta_c = \omega_0/\Omega_{ce}$. Right: iso-contours of constant growth rate $\gamma$ in the $k_\perp - k_\parallel$ plane obtained by solving the full dispersion relation for the secondary modes. \change{The modes with relatively high $k_\perp$ are related to ECDI, while the modes with smaller $k_\perp$ near the critical angle correspond to the electrostatic whistlers in the frame of reference oscillating with the cold electrons.}}
\label{fig:low-A-k-spectrum}
\end{figure}

The expectations summarized above are confirmed by the analysis of the spectrum of ion density perturbations in the low-anisotropy simulation, which demonstrates excitation of both quasi-perpendicular ECDI-like instabilities and the instabilities near the critical angle corresponding to the driver frequencies, as shown in the left panel of Fig.~\ref{fig:low-A-k-spectrum}. \change{Both instabilities couple to cold ions, which is possible because they are Doppler shifted in the ion frame of reference}. The right panel of Fig.~\ref{fig:low-A-k-spectrum} shows iso-contours of constant growth rates $\gamma$ for the two indicated values  obtained by numerically solving the full dispersion relation for the secondary modes (obtained by combining Eqs.~\ref{eq:electrons_oblique} and~\ref{eq:ion_response_transformed}). We observe that the theoretical analysis correctly predicts the wavenumbers of the instabilities observed in the simulation and yields values for the growth rate consistent with those measured in the simulation (not shown). We note that in contrast to the quasi-perpendicular modes with $k_\perp \rho_e^C \gtrsim 1$, oblique modes appear at $ k_\perp \sim 40 d_e^{-1} \sim 0.3 (\rho_e^C)^{-1}$. \change{In the frame of reference oscillating with cold electrons, these modes correspond to electrostatic whistlers driven unstable by coupling to Doppler-shifted ion response.}

\begin{figure}[h]
\centering
\includegraphics[width=6in]{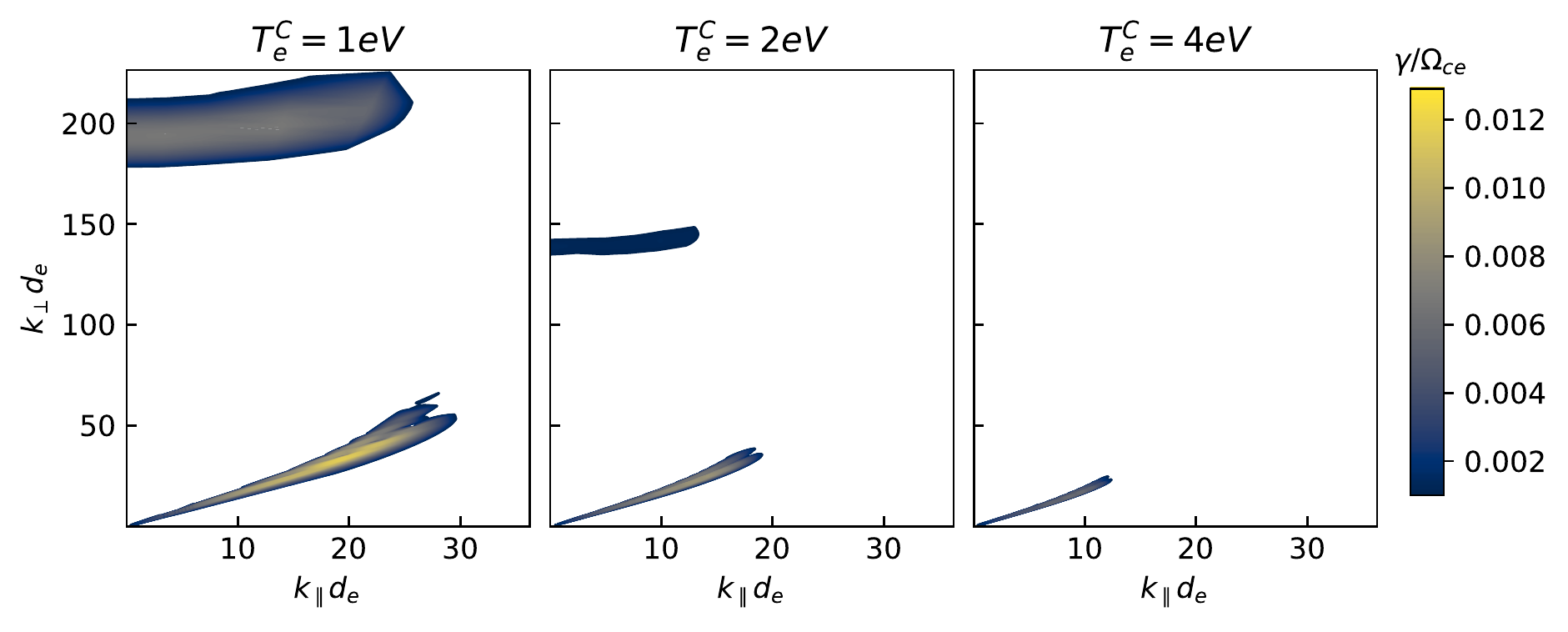}
\caption{\change{Growth rate of the secondary modes for 3 values of the cold electron temperature. The parameters of the driver and the density of the cold electrons are fixed and correspond to the lower-anisotropy simulation with $\delta B/B_0 \sim 5\times 10^{-4}$.}} 
\label{fig:linear_T_scan}
\end{figure}

\change{It is instructive to examine how the properties of the secondary modes change with the properties of the cold population. An example of such an analysis is presented in Fig.~\ref{fig:linear_T_scan}, which shows the variation of the growth rate as a function of the temperature of the cold populations $T_e^C$. The other parameters were chosen to correspond to the lower-anisotropy case. The peak growth rate for the quasi-perpendicular ECDI-like modes moves towards lower $k_\perp$ with increasing $T_e^C$, consistent with the expectation that the modes are characterized by $k_\perp \rho_e^C \sim 1$. The peak growth rate of the oblique whistler-like modes also moves to lower $k$ with increasing temperature of the cold population, while the angle corresponding to the maximum growth remains close to $\theta_c$. Note also that the peak growth rate decreases with increasing cold electron temperature, consistent with the expectation that the key parameter controlling the secondary drift instabilities is the amplitude of the induced cold-electron flow relative to the cold electron thermal velocity. Finally, we remark that the presented analysis is within the electrostatic approximation, which can be expected to hold for modes with $k d_e \gg 1$. Electromagnetic effects may modify the behavior at lower values of $k d_e \sim 1 $.}

\begin{figure}[h]
\centering
\includegraphics[width=5in]{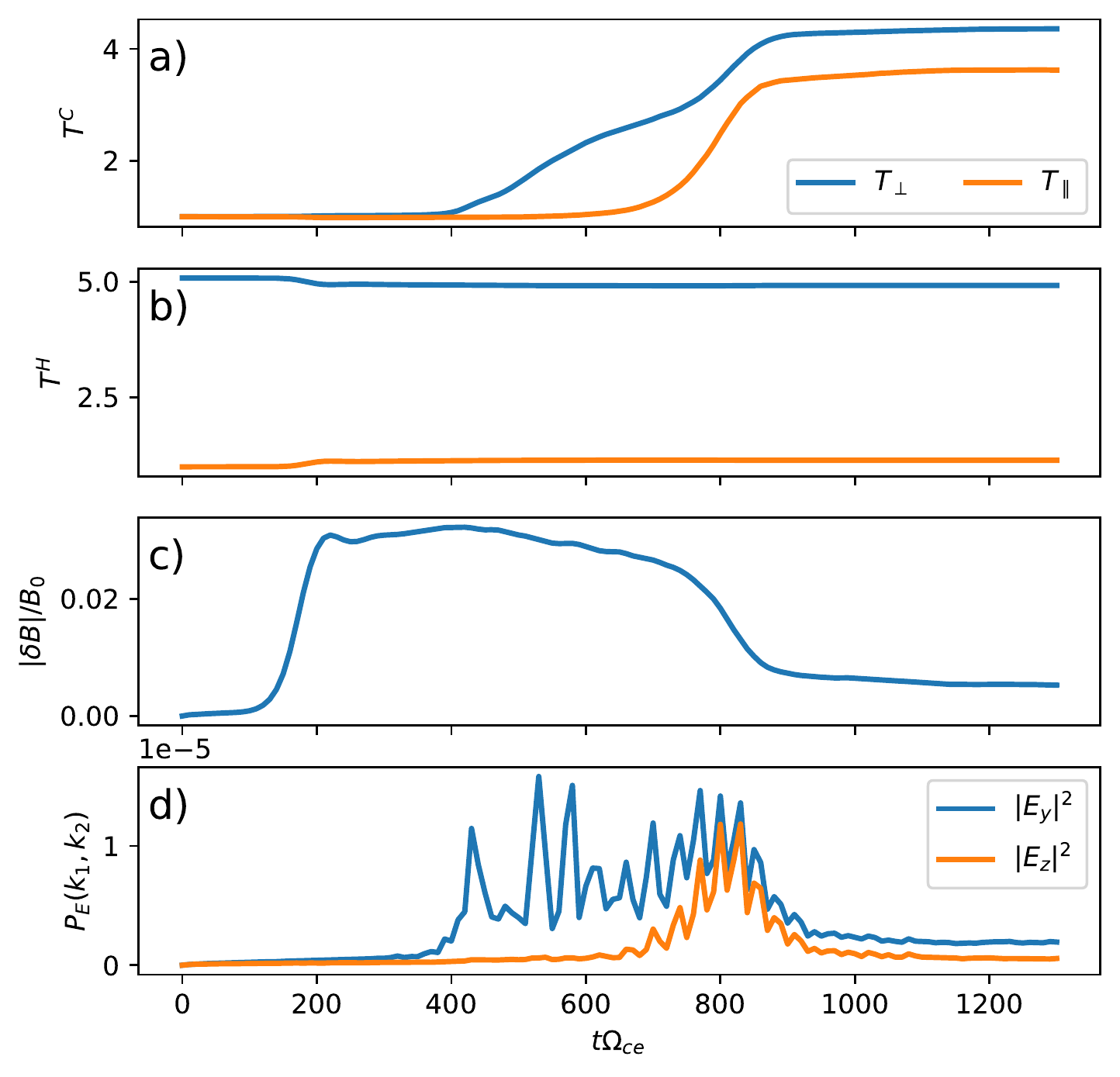}
\caption{Results of the 3D simulation, in the format similar to Fig.~\ref{fig:mainresult5}: \change{(a) the parallel and perpendicular temperatures of the cold electron population, (b) the parallel and perpendicular temperatures of the hot electron population, (c) the amplitude of magnetic field fluctuations. } Panel (d) shows power in short-wavelength fluctuations in both $E_y$ (characteristic of predominately perpendicular modes) and $E_z$ (characteristic of the oblique modes). The strongest decay of the primary mode is correlated with onset of oblique instabilities.} 
\label{fig:3D}
\end{figure}

In order to further illustrate significance of the processes described here, Fig.~\ref{fig:3D} shows results of a 3D simulation with parameters corresponding to the high-anisotropy 2D case discussed above. In 3D geometry, multiple secondary modes at different orientations are allowed to develop. We observe that quasi-perpendicular modes, characterized by high amplitude of $E_x$ and $E_y$ fluctuations are excited first and lead to moderate damping of the primary whistler and perpendicular heating of the cold background. However, excitation of the oblique fluctuations, which are characterized by strong $E_z$ fluctuations, leads to much stronger damping of the primary mode and fast increase of both parallel and perpendicular temperatures of the background electrons. For the parameters considered, the  amplitude of the primary whistler is reduced by approximately a factor of 6 relative to the peak value. 

\section{Conclusions and Discussion}
 To summarize, we have demonstrated that the presence of cold electron population introduces coupling of the whistler modes to short-wavelength, oblique, electrostatic instabilities. This coupling is driven by a relative drift between \change{the cold ion and cold} electron populations induced by the fluctuating electric field of the whistler waves. For the parameters considered in this study, two of the most prominent instabilities are related to the Electron Cyclotron Drift Instability (ECDI)~\cite{Forslund1970,Forslund1972} and the electrostatic whistlers. Both of these short-wavelength instabilities lead to damping of the primary whistler mode and heating of the background cold population. For the parameters considered, the ECDI-type instabilities appear first in the simulations. They lead predominantly to perpendicular heating of the cold population and a modest damping of the primary whistler modes. Oblique electrostatic whistlers appear at later times, but lead to a much faster decay of the primary whistler and stronger isotropic heating of the cold background population due to relatively large fluctuations of the parallel electric field. \change{It is is interesting to note that oblique electrostatic whistler modes can also be driven unstable by anisotropy of the cold electrons~\cite{Hashimoto1981}. Since the ECDI-like modes predominantly heat the cold electrons in the perpendicular direction, they may induce growth of the oblique whistler if sufficient anisotropy of the cold electrons is generated. Finally, we note that proper description of the discussed instabilities requires challenging multi-dimensional simulations that properly resolve scales associated with the cold electron population (e.g. the cold electron gyroraidus), which might explain why they appear to have been missed in the previous investigations.}
  
The processes discussed in this paper may, in principle, have important impact on the propagation of whistler waves in the Earth's magnetosphere. For example, in the presence of a cold electron population, whistler waves are limited to lower amplitudes $\delta B$ than without it. Since the efficiency of wave-particle interactions  generally scales with $\delta B$ (for instance, as $\delta B^2$ in quasi-linear theory), the cold populations could (indirectly) play a \change{significant} role in determining the dynamics of the environment. The presented results thus highlight the significance of the cold plasma populations, which are often viewed as “passive”, simply providing the bulk plasma density. The properties of the cold plasma are relatively poorly understood due to the difficulties associated with direct spacecraft measurements~\cite{GLD-cold}. In the scenario proposed here, the cold populations play an active role, which emphasizes the need for the better understanding of such “hidden” magnetospheric populations. 

\change{Unfortunately, the lack of accurate measurements of the cold electron and ion populations (in particular the temperature but more generally the energy distribution) in the relevant regions of the magnetosphere makes unequivocal identification of the processes described here challenging. The main observational signature available from current measurements would likely be observation of high-frequency electrostatic oscillations (with frequencies up to and exceeding electron cyclotron frequency) in the presence of whistler waves in the regions where the density of cold plasma is significant. 
Strictly speaking, such oscillations do not correspond to “normal” plasma waves in stationary uniform plasma. Instead, they are eigenmodes of a nonlinear state that essentially depends on the presence of an oscillating electric field associated with the primary whistler wave. When viewed in a frame of reference that is oscillating with the cold electrons, these modes correspond to the intersection of a Doppler-shifted ion response with the dispersion relation of electrostatic whistlers (for the oblique modes) or that of the electron Bernstein modes (for the nearly perpendicular modes). In the stationary frame of reference, such oscillations will appear in one or more frequency bands separated by the harmonics of the primary whistler mode. 
It is interesting to note that the statistical studies of chorus waves from various spacecraft missions~\citep{hayakawa1984,haque2010chorus,li2011chorus,agapitov2012corr,li2013,li2016,teng2019} indicate that oblique orientation near the resonance cone is commonly seen, even though orientation nearly parallel to the direction of the local magnetic field is the most probable. It is possible that coupling to the cold plasma contributes to the generation of oblique chorus waves, although many other generation mechanisms have also been proposed~\citep{li2016,Fu2017}.

}

It should also be emphasized that the processes described in this paper affect any whistler waves of sufficient amplitude, regardless of their origin. They could therefore affect artificially injected waves, and as such must be considered in the analysis of radiation belt remediation schemes based on whistler waves artificially injected in the environment to induce \change{particle} losses and reduce harmful fluxes of relativistic electrons to levels that are tolerable for our space infrastructure~\cite{ganguli2015,carlsten19}. Furthermore, the discussed processes enable whistler modes to heat cold electrons. In the plasmasphere, where whistler-mode hiss waves are present, this could provide an additional heat source for the cold plasma that might help explaining why models of the plasmasphere are consistently underestimating the temperature with respect to \change{available} observations~\cite{comfort96,Bezrukikh2006,gallagher2011}.

We conclude by briefly discussing the limitations of the presented analysis. Our results highlight the existence of a class of nonlinear processes that may affect the dynamics of whistler waves in the magnetosphere. Whether these processes play an important role in any given scenario will be determined to a large degree by the proprieties of the cold populations, which are poorly quantified at present. For the parameters chosen in the simulations, the most unstable primary whistler modes are field-aligned. It is well known that for values of $\beta_{||e}^H$ below a certain threshold, the maximum growth rate corresponds to oblique waves~\cite{gary12}, although the presence of cold populations lowers this threshold value.  The relative importance of the secondary instabilities and Landau damping due to parallel electric field for oblique whistlers will need to be explored. Additionally, our simulations are local, focusing on a relatively small domains with periodic boundary conditions. In the real situation, the waves can propagate outside the source region and might return to it (possibly with different amplitude) only if they are reflected back at higher latitudes. \change{With the exception of one case, the simulations are performed in two spatial dimensions, with magnetic field in the plane of the simulation. Such a configuration suppresses nonlinear scattering of whistlers~\cite{Ganguli2010}, which could be an important effect in low-$\beta$ plasmas. Further}, we have considered a uniform background magnetic field. For chorus waves, it is well known that a non-uniform magnetic field is important as it might lead to frequency chirping and the formation of rising or falling chorus elements~\cite{Omura2012}. \change{While the results presented here provide clear evidence for a new nonlinear mechanism affecting whistler waves, it will be} important to assess the role played by a non-uniform magnetic field as well as the relative significance of other nonlinear mechanisms involving whistlers that have been previously identified in the literature.

\acknowledgments
 We thank Joe Borovsky, Lauren Blum, Craig Kletzing, Lynn Wilson III, and Cynthia Cattell for stimulating discussions. VR was supported by NSF grant 1707275. GLD was supported by the Laboratory Directed Research and Development program at Los Alamos National Laboratory (LANL) under project  20200073DR. LANL is operated by Triad National Security, LLC, for the National Nuclear Security Administration of U.S. Department of Energy (DOE) (Contract No. 89233218CNA000001). Computational resources supporting this work were provided by the NASA High-End Computing (HEC) Program through the NASA Advanced Supercomputing (NAS) Division at Ames Research Center and by the National Energy Research Scientific Computing Center (NERSC), a U.S. Department of Energy Office of Science User Facility operated under Contract No. DE-AC02-05CH11231.

\section*{Data Availability}
The data that support the findings of this study are available from the corresponding author upon reasonable request.


%
%

\bibliography{whistler,chorus_lit}

%
%
%
%
%

\end{document}